\documentclass{article}
\usepackage{spconf,amsmath,graphicx}

\usepackage{enumitem}
\setlist{nosep, leftmargin=14pt}
\usepackage{booktabs}

\usepackage{mwe} % to get dummy images
\usepackage{svg}
\usepackage{url}
\usepackage{eso-pic}
\newcommand{\ieeecopyright}{%
\AddToShipoutPictureFG*{%
  \AtPageLowerLeft{%
    \raisebox{10mm}{%
      \makebox[\paperwidth]{%
        \parbox{0.83\paperwidth}{%
          \centering
          \footnotesize
          © 2026 IEEE. Personal use of this material is permitted.
          Permission from IEEE must be obtained for all other uses,
          in any current or future media, including reprinting/republishing
          this material for advertising or promotional purposes, creating
          new collective works, for resale or redistribution to servers or
          lists, or reuse of any copyrighted component of this work in
          other works.
        }%
      }%
    }%
  }%
}}

\title{Log Focal Frequency Loss for Bioimage Restoration}
\name{%
\begin{tabular}{@{}c@{}}
Xingjian Zhang$^{\star \dagger}$ \qquad 
Claire Leclech$^{\star}$ \qquad 
Louison Blivet-Bailly$^{\star}$ \\ 
Abdul I. Barakat$^{\star}$ \qquad 
Elsa D. Angelini$^{\dagger}$
\end{tabular}}
\address{$^{\star}$ LadHyX, CNRS, École polytechnique, Institut Polytechnique de Paris, France \\ $^{\dagger}$ LTCI, Télécom Paris, Institut Polytechnique de Paris, France 
    }

\begin{document}
%\ninept

%
\maketitle
\ieeecopyright

\begin{abstract}
Image restoration of biological structures in microscopy poses unique challenges for preserving fine textures and sharp edges. While recent GAN-based image restoration formulations have introduced frequency-domain losses for natural images, microscopy images pose distinct challenges with large dynamic ranges and sparse but critical structures with spatially-variable contrast. Inspired by the principle of logarithmic perception in human vision, we propose a log focal frequency loss (LFFL) tailored for microscopy restoration. This loss combines adaptive spectral weighting from log-space differences with log-dampened error measurement, ensuring balanced reconstruction across all frequency bands while preserving both structural coherence and fine details. We tested our GAN-based framework on two use-cases with real ground-truths: deblurring of fluorescence images of cell nuclei on microgroove substrates and denoising of zebrafish embryo images from the FMD dataset. Compared to training with only spatial-domain losses and with existing frequency-domain losses, our method achieves improvements across several quality metrics. Code is available at \url{github.com/xjzhaang/log-focal-frequency-loss}.
%While developed for microscopy, the method may generalize to other imaging domains where high-frequency detail preservation is critical. %Code will be made available upon acceptance.
\end{abstract}

\begin{keywords}
image restoration, GAN, fluorescence microscopy, frequency domain loss, deblurring, denoising
\end{keywords}

\section{Introduction}
\label{sec:intro}

%Fluorescence microscopy is fundamental to research, diagnostics, and drug discovery. Obtaining sharp, high-quality images is essential for accurate analysis of biological structures and their dynamics. However, 
Fluorescence microscopy images are frequently blurry, out-of-focus, or noisy due to optical limitations, specimen thickness, or suboptimal imaging conditions.
Deep learning-based image denoising and deblurring methods have become powerful tools for enhancing image quality, yet they face a fundamental limitation: neural networks trained with gradient descent exhibit spectral bias \cite{rahaman2019on}—an inherent bias toward learning low-frequency components while struggling to capture high-frequency details.
Spatial losses fail to counteract this limitation as they uniformly average errors across pixels, diluting gradients of high-frequency features. While current frequency-domain formulations explicitly target spectral content, they typically rely on raw frequency amplitudes \cite{fuoli2021, jiang2021} where larger low-frequency amplitudes disproportionately influence the loss.
This limitation is particularly problematic for fluorescence microscopy images which often contain only sparse bright structures against low-intensity backgrounds. Meaningful information primarily resides in high-frequency components representing edges and fine textures, but the foreground-background imbalance compounds spectral bias, rendering high-frequency restoration even more challenging. To address this gap, we propose a log focal frequency loss that re-balances learning across frequency scales into a standard GAN-based framework. We introduce a logarithmic focal weight and a log-dampened error to equalize the relative contributions of different frequency components. 
%Integrated into a standard GAN-based framework alongside pixel-level, perceptual, and adversarial losses, the proposed GAN enhances restoration quality and better preserves the fine structural details essential for accurate biological analysis.

\subsection{Related Work}
\label{ssec:relwork}

Classical image restoration losses, such as pixel-wise L1 or L2 norms, prioritize overall intensity accuracy and tend to suppress fine details such as edges and textures \cite{Zhao2017}. Perceptual losses based on deep feature activations improve semantic fidelity \cite{Johnson2016}, yet in microscopy images they remain biased toward background regions, as spatial averaging dilutes the contribution of localized biologically-relevant structures.

Recent frequency-domain approaches for image reconstruction include neural network architectures that explicitly process spectral representations \cite{chen2024, Duan2025} and loss formulations that operate in the frequency domain \cite{fuoli2021, jiang2021}, often to address the inherent spectral bias of neural networks toward learning low-frequency components \cite{rahaman2019on}.
Specific frequency-domain losses have decomposed the problem in various ways. Fuoli et al. \cite{fuoli2021} proposed the Fourier Space Loss (FSL), which separately computes L1 losses on amplitude and phase components of the Fourier transform. Focal Frequency Loss (FFL) \cite{jiang2021} took a different approach, introducing adaptive weighting based on frequency-wise errors to dynamically emphasize harder-to-learn frequencies. 
However, both approaches do not account for the fact that the amplitudes of low-frequency components can be orders of magnitude larger than those of high frequencies, causing their losses to be disproportionately weighted toward low-frequency regions. 
%This issue is particularly relevant for microscopy, where meaningful information often resides in the high-frequency domain representing fine structural details.

\section{Methods}
\label{sec:method}

\subsection{Log Focal Frequency Loss}
\label{ssec:lsfp}

We propose a weighted log frequency loss tailored to microscopy images, building upon the FFL framework introduced in \cite{jiang2021} for natural image reconstruction and synthesis. We modify this framework by introducing logarithmic scaling at two stages: (1) computing adaptive weights from log differences of real and imaginary components, and (2) measuring reconstruction error as the log-dampened magnitude of complex frequency differences. This differs from existing frequency losses \cite{fuoli2021, jiang2021} that compute distances using the raw magnitudes of complex differences or separate amplitude/phase components. 

For training, we work with pairs of low-resolution (LR) and high-resolution (HR) images. 
We aim to infer the HR image from its LR observation.
We apply the 2D discrete fast Fourier transform (FFT) with orthonormal normalization to the predicted and ground-truth representations, converting spatial representations $f(x,y)$ into their frequency domain counterparts $\mathcal{F}(u,v)$. 
%Depending on the task, these representations are either raw image intensities (for denoising) or perceptual features from intermediate VGG layers (for deblurring).
For each frequency component at position $(u,v)$, we decompose its complex Fourier value into real (Re) and imaginary (Im) parts. Following the FFL distance formulation of \cite{jiang2021}, this decomposition preserves both amplitude and phase information. We first compute log-space differences between the ground-truth $\mathcal{F}_{\text{target}}$ and the predicted image $\mathcal{F}_{\text{pred}}$ as:
\begin{equation}
\begin{split}
\Delta_{\text{Re}}(u,v) &= \log|\text{Re}(\mathcal{F}_{\text{pred}}) + \epsilon | - \log|\text{Re}(\mathcal{F}_{\text{target}}) + \epsilon| \\
\Delta_{\text{Im}}(u,v) &= \log|\text{Im}(\mathcal{F}_{\text{pred}}) + \epsilon| - \log|\text{Im}(\mathcal{F}_{\text{target}}) + \epsilon|
\end{split}
\end{equation}
\noindent where $\epsilon = 10^{-8}$ is used for numerical stability.

These log-difference maps are combined into an adaptive weight map that emphasizes frequencies with larger discrepancies as:
\begin{equation}
w_{\text{rel}}(u,v) = \left(\sqrt{\Delta_{\text{Re}}^2 + \Delta_{\text{Im}}^2}\right)
\label{ssec:eq2}
\end{equation}

To compute the loss, we also measure the log-dampened reconstruction error:
% \begin{equation}
% \varepsilon_{\text{log}}(u,v) = \log\left(|\mathcal{F}_{\text{pred}}(u,v) - \mathcal{F}_{\text{target}}(u,v)| + 1\right)
% \end{equation}
\begin{equation}
D_{\text{log}}(u,v) = \log\left(|\mathcal{F}_{\text{pred}}(u,v) - \mathcal{F}_{\text{target}}(u,v)| + 1\right)
\end{equation}

Logarithmic dampening ensures that errors are evaluated on a compressed scale across the wide dynamic range of frequency amplitudes. The gradient with respect to the predicted frequency coefficients satisfies:
\begin{equation}
\left|\frac{\partial D_{\text{log}}}{\partial \mathcal{F}_{\text{pred}}}\right| \propto \frac{1}{|\mathcal{F}_{\text{pred}} - \mathcal{F}_{\text{target}}| + 1}
\label{ssec:grad}
\end{equation}
    This inverse relationship means that smaller errors receive proportionally stronger gradient signals than larger ones, preventing high-amplitude background-based reconstruction errors from overwhelming the optimization. 

Our proposed Log Focal Frequency Loss (LFFL) is defined as: 
\begin{equation}
\mathcal{L}_{\text{LFFL}} = \frac{1}{MN} \sum_{u=0}^{M-1} \sum_{v=0}^{N-1} w_{\text{rel}}(u,v)^\alpha \cdot D_{\text{log}}(u,v)
\end{equation}
where $\alpha>0$ is a ``focal" scaling factor that controls emphasis on relatively harder frequencies. 

When manipulating Fourier coefficients, the DC component (at $(u,v)=(0,0)$ and equal to the average value of the image) requires specific care. In our implementation, we zero out the DC components of $\mathcal{F}_{\text{pred}}$ and $\mathcal{F}_{\text{target}}$ for computing the weight map $w_{\text{rel}}$, which effectively excludes the DC component from contributing to the loss.

\subsection{Overall Training Loss}
\label{ssec:overall_loss}
In our GAN architecture, the Generator training loss combines standard spatial losses (pixel-level, perceptual, adversarial) with our proposed LFFL term as:

\begin{equation}
\mathcal{L}_{\text{total}} = \lambda_{1} \mathcal{L}_{\text{pixel}} + \lambda_{2} \mathcal{L}_{\text{perc}} + \lambda_{3} \mathcal{L}_{\text{adv}} + \lambda_{4} \mathcal{L}_{\text{LFFL}}
\end{equation}

\noindent where $\lambda_{1}$, $\lambda_{2}$, $\lambda_{3}$, and $\lambda_{4}$ constitute weight factors for the L1 pixel-wise loss, a VGG-based perceptual loss \cite{Johnson2016}, an adversarial loss, and our proposed LFFL, respectively. The adversarial loss follows the relativistic average GAN formulation \cite{JolicoeurMartineau2018TheRD}. Specific weight factor values are provided in Section~\ref{ssec:implementation}.

\subsection{Selection of Frequency Content for LFFL}
We compute LFFL on the FFT of different input spatial representations depending on the restoration task. For \textbf{deblurring}, we apply LFFL to the perceptual features extracted from VGG-19 \cite{VGG} layer \texttt{conv1\_2}. Early VGG layers capture information such as edges and textures, which aligns well with the restoration goal. For \textbf{denoising}, we compute LFFL directly on image intensities to preserve fidelity to the raw signal, ensuring genuine high-frequency structural details are retained.

% Unlike existing methods that compute frequency losses directly on image FFTs, we apply LFL to perceptual features extracted from early VGG layer \cite{VGG} for both deblurring and denoising tasks. Early VGG layers capture essential structural information such as edges and textures, while the network itself inherently provides a denoising effect, making these features ideal for frequency-domain supervision in microscopy applications.

% \textbf{Task-specific frequency representations.}  We compute the LFL either on image intensities (for denoising) or on perceptual feature maps extracted from intermediate VGG layers \cite{VGG} (for deblurring). Using perceptual features for deblurring offers a key advantage: they naturally suppress low-level pixel noise and acquisition artifacts while emphasizing biologically meaningful structures, aligning frequency-domain supervision with microscopy content rather than measurement noise. Conversely, computing LFL directly on image intensities for denoising preserves fidelity to the raw signal, ensuring genuine high-frequency structural details are retained while stochastic noise is suppressed. %We do this for deblurring not for denoising maybe - to check

\begin{table*}[!hbtp]
\centering
\caption{Quantitative comparison of spatial and frequency-domain losses on fluorescence microscopy datasets. Bold indicates best mean result per metric. $^*$ indicates statistical significance (p $<$ 0.001, paired t-test) compared to the best baseline metric.}
\vspace{1mm}
\label{tab:main_results}
\resizebox{\textwidth}{!}{
\begin{tabular}{c *{6}{c} c *{6}{c}}
\toprule
& \multicolumn{6}{c}{\textbf{Out-of-focus deformed nuclei (N=877)}}
& &
\multicolumn{6}{c}{\textbf{Noisy zebrafish embryo (N=600)}} \\
% \cmidrule(lr){2-7} \cmidrule(l){8-13}
\cmidrule(lr){2-7} \cmidrule(lr){9-14}
Method & PSNR$\uparrow$ & SSIM$\uparrow$ & LPIPS$\downarrow$ & FID$\downarrow$ & FSIM$\uparrow$ & GMSD$\downarrow$
& & 
PSNR$\uparrow$ & SSIM$\uparrow$ & LPIPS$\downarrow$ & FID$\downarrow$ & FSIM$\uparrow$ & GMSD$\downarrow$ \\

\midrule
Spatial only & 36.297\phantom{$^*$} & 0.876\phantom{$^*$} & 0.0175\phantom{$^*$} & 30.146\phantom{$^*$} & 0.929\phantom{$^*$} & 0.0727\phantom{$^*$} 
&  & 33.675\phantom{$^*$} & 0.897\phantom{$^*$} & 0.245\phantom{$^*$} & 93.434\phantom{$^*$} & 0.919\phantom{$^*$} & 0.0503\phantom{$^*$} \\

FFL \cite{jiang2021} & 35.934\phantom{$^*$}  & 0.872\phantom{$^*$}  & 0.0177\phantom{$^*$}  & 31.002\phantom{$^*$}  & 0.929\phantom{$^*$}  & 0.0731\phantom{$^*$}  
&  & 33.793\phantom{$^*$}  & 0.900\phantom{$^*$}  & \textbf{0.241}\phantom{$^*$}  & 93.559\phantom{$^*$}  & 0.920\phantom{$^*$}  & 0.0492\phantom{$^*$}  \\

FSL \cite{fuoli2021} & 35.748\phantom{$^*$} & 0.866\phantom{$^*$}  & 0.0190\phantom{$^*$}  & 32.449\phantom{$^*$}  & 0.924\phantom{$^*$}  & 0.0780\phantom{$^*$}  
&  & 33.737\phantom{$^*$}  & 0.899\phantom{$^*$}  & 0.248\phantom{$^*$}  & \textbf{91.077}\phantom{$^*$}  & 0.920\phantom{$^*$}  & 0.0501\phantom{$^*$}  \\

LFFL ($\alpha=0.5$)  & 36.184$^{\smash{*}}$ & $0.885^*$ & \textbf{0.0157}$^*$ & 29.345\phantom{$^*$}  & \textbf{0.937}$^*$ & 0.0699$^*$
%& 33.778  & 0.900  & \textbf{0.239} & 90.264 & 0.919 & 0.0500  \\
&  & 34.118{$^*$} & \textbf{0.911}{$^*$} & 0.266\phantom{$^*$}  & 98.848\phantom{$^*$}  & \textbf{0.921}{$^*$} & \textbf{0.0488}{$^*$} \\

LFFL ($\alpha=1$)  & $36.674^*$ & $0.883^*$ & 0.0173\phantom{$^*$}  & \textbf{28.169}\phantom{$^*$}  & 0.933$^*$ & 0.0706$^*$ 
&  &
%\textbf{33.890} & \textbf{0.902} & 0.242 & 93.475 & 0.919 & 0.0487 \\
34.142{$^*$} & 0.910{$^*$} & 0.266\phantom{$^*$}  & 95.958\phantom{$^*$}  & 0.917\phantom{$^*$}  & 0.0505\phantom{$^*$}  \\

LFFL ($\alpha=2$) & \textbf{36.856}$^*$ & \textbf{0.889}$^*$ & 0.0166$^*$ & 37.111\phantom{$^*$}  & 0.936$^*$ & \textbf{0.0679}$^*$
&  & \textbf{34.163}{$^*$} & \textbf{0.911}{$^*$} & 0.274\phantom{$^*$}  & 95.926\phantom{$^*$}  & 0.914\phantom{$^*$}  & 0.0515\phantom{$^*$}  \\

\bottomrule
\end{tabular}
}
\end{table*}

\section{Experiments}
\label{sec:exp}

\subsection{Datasets}
\label{ssec:datasets}

We evaluate our method on two fluorescence microscopy datasets with varying acquisition characteristics and degradation types, aiming to illustrate the capacities of our proposed LFFL lose in deblurring and denoising tasks:

% \item \textbf{Synthetically degraded cellular dataset (in-house):} Fluorescence microscopy images of human umbilical vein endothelial cells (HUVECs) staining nuclei (DAPI), intermediate filaments (Vimentin), cell junctions (VE-Cadherin), and actin filaments (Phalloidin). [X] images (2048$\times$2044 pixels) at 0.33$\mu$m/pixel acquired using a 20X objective. Synthetic degradations simulate blur and noise via convolution with physically realistic confocal PSFs followed by Poisson-Gaussian noise \cite{Li2024}.

% For the in-house datasets, paired images are captured sequentially on the same fields of view using [microscope model] with [objective]. The $\pm 5~\mu$m defocus range represents typical imaging depths in [application].

\begin{itemize}
\item \textbf{Deblurring: Out-of-focus deformed nuclei dataset (in-house).} This use-case focuses on fluorescence microscopy images of fibroblast cells stained for lamin A/C to delineate the nuclei. The cells were cultured on microgroove substrates to induce 3D nuclear deformations \cite{Leclech2025}. The dataset contains 945 image pairs (of size 2,044$\times$2,048 pixels) acquired at a resolution of 0.33 $\mu$m/pixel using a 20X microscope objective. Low-resolution images were acquired with a defocus of $\pm 5~\mu$m from the optimal focal plane.

\item \textbf{Denoising: Noisy zebrafish embryo dataset (open-access).} This use-case focuses on fluorescence microscopy images of EGFP-labeled zebrafish embryos from the FMD dataset \cite{zhang2018poisson}. It contains 1,000 image pairs (of size 512$\times$512 pixels) with real noisy observations and corresponding averaged ground-truth denoised images.

\end{itemize}

We split each dataset into training and test sets at the highest hierarchical level (experiment or FOV) to prevent data leakage, targeting a 40\%/60\% ratio. 

\subsection{Implementation Details}
\label{ssec:implementation}
We use a 16-layer RRDB Generator from ESRGAN \cite{Wang2019} and a PatchGAN Discriminator \cite{Isola2017}. 
We extract patches of size $256 \times 256$ pixels from regions of interest and normalize patch intensity values to [0,1] using the bit depth for the maximal value. This yields 1,440 patches from the nuclei dataset and 1,000 patches from the zebrafish dataset.
Loss weights were empirically set to: $\lambda_1 = 0.1$ for deblurring, $\lambda_1 = 1$ for denoising, $\lambda_2 = 1$, and $\lambda_3 = 0.005$. 
The weight factor $\lambda_4$ for frequency-domain loss $\mathcal{L}_{\text{LFFL}}$ was calibrated per model to achieve comparable initial gradient magnitudes and remained fixed during training. For our three models with $\alpha \in \{0.5, 1, 2\}$, we set $\lambda_4 \in \{2, \sqrt{2}, \sqrt[4]{2}\}$ for deblurring and $\lambda_4 \in \{4, 2, \sqrt{2}\}$ for denoising, following an inverse relationship with $\alpha$.

The perceptual loss uses \texttt{conv2\_2} layer features from a VGG-19 pretrained on ImageNet. 
We use AdamW optimizer with $\beta_1 = 0.9$, $\beta_2 = 0.999$, learning rates of $10^{-4}$ (Generator) and $10^{-5}$ (Discriminator), with dynamic Discriminator scheduling for training stability. Models were trained for 200 epochs with batch size 4.

\subsection{Evaluation Metrics}
We evaluate PSNR, SSIM, FSIM \cite{LinZhang2011}, and GMSD \cite{Xue2014} at the patch level for the zebrafish dataset and at the nuclear level for the nuclei dataset, where we extract nuclear bounding-boxes using masks from Cellpose \cite{Stringer2020}. LPIPS-VGG \cite{Zhang2018} is reported at the patch level and FID \cite{fid} at the dataset level.

\subsection{Benchmark Methods}
We compare LFFL against three baselines: spatial baseline ($\mathcal{L}_{\text{pixel}}$, $\mathcal{L}_{\text{perc}}$, and $\mathcal{L}_{\text{adv}}$ only), FFL~\cite{jiang2021}, and FSL~\cite{fuoli2021} excluding their Fourier Discriminator to isolate loss contributions\footnote{FSL was originally designed to work with a Fourier-domain Discriminator. We evaluate its loss formulation in isolation only to contrast weighting behavior, not to assess the full model's performance.}. 
All methods use identical training/test sets, architectures and training configurations. 
FFL and FSL are computed on image FFTs as described in their respective papers, with weights calibrated to match LFFL's initial gradient magnitudes. 

\section{Results}
\label{ssec:results}

\begin{figure*}[t]
\centering
\includegraphics[width=\textwidth]{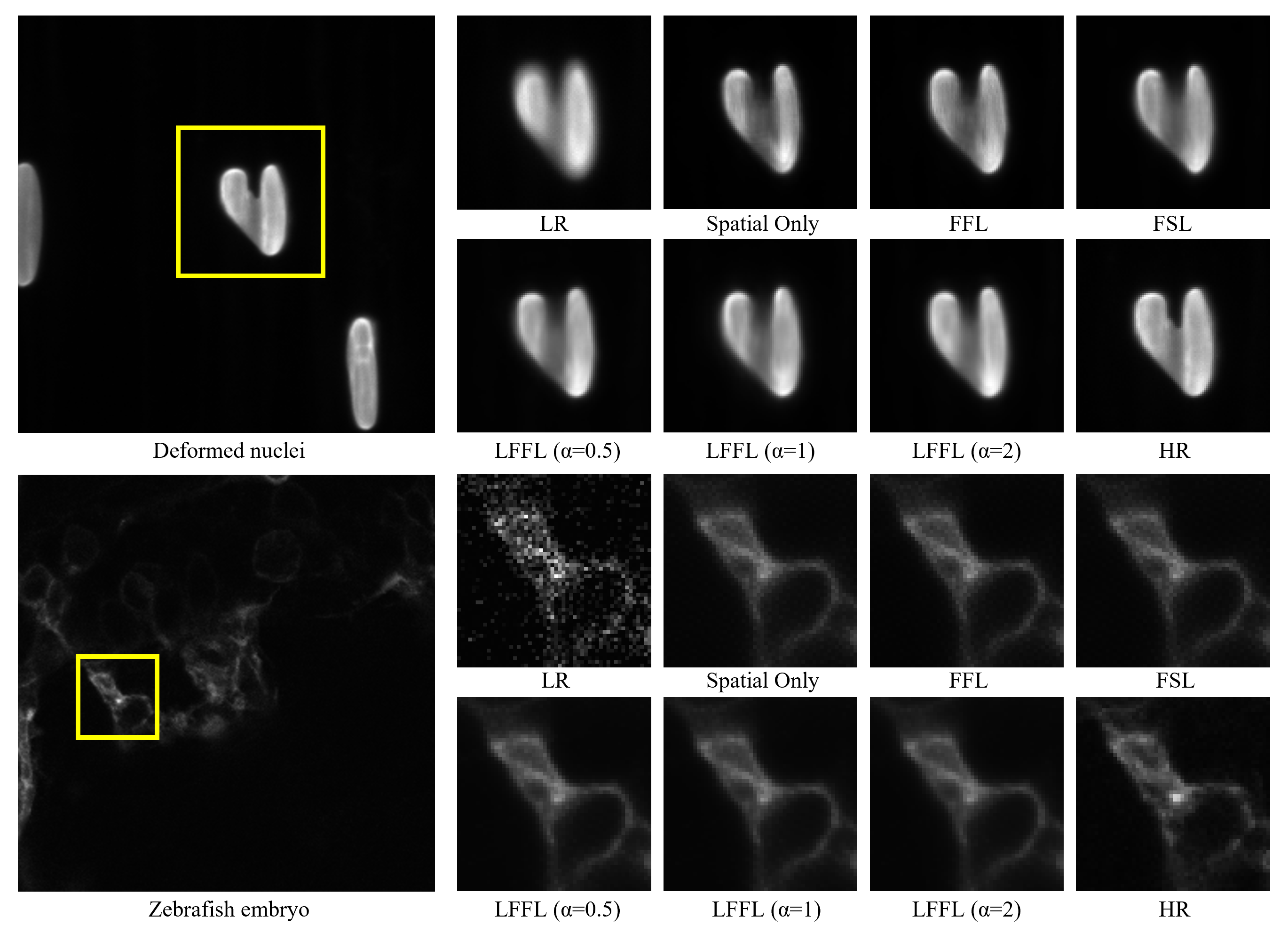}
\caption{Visual comparison of deblurring (top) and denoising (bottom) results with magnified regions from yellow boxes. %For deblurring, LFFL recovers more faithful subcellular textures compared to baselines (LR: low-resolution input, Spatial Only, FFL, FSL, HR: high-resolution ground truth), which introduce faulty details and artifacts. For denoising, LFFL achieves superior noise suppression while preserving fine morphological details. 
}
\label{fig:qualitative_results}
\end{figure*}

Results for deblurring and denoising are reported in Table~\ref{tab:main_results}. The proposed LFFL consistently outperforms spatial and frequency-domain losses on the deblurring task across all metrics. For denoising, where pixel-level fidelity is more critical, LFFL achieves gains in PSNR and SSIM while maintaining competitive perceptual performance. The more modest improvements for denoising may be attributed to imperfect ground-truth images, which contain residual noise that limits the effectiveness of frequency-domain supervision.

Figure~\ref{fig:qualitative_results} shows some visual restoration results. For deblurring, LFFL recovers cleaner nuclear boundaries with fewer artifacts and more faithful subcellular textures while maintaining overall structural coherence. Other losses tend to introduce spurious details and boundary irregularities in defocused regions. For denoising, LFFL achieves superior noise suppression with smoother intensity transitions, whereas baseline methods exhibit residual noise manifesting as pixelation and graininess artifacts.

 We evaluate the scaling factor $\alpha$  in Eq.~\ref{ssec:eq2} across values of 0.5, 1, and 2. For both tasks, $\alpha=0.5$ yields better perceptual metrics while $\alpha=2$ achieves higher PSNR/SSIM, reflecting a trade-off between amplifying large relative errors ($\alpha=2$, aggressive hard-frequency focus) and compressing error magnitudes ($\alpha=0.5$, reduced sensitivity to large errors).

\section{Discussion}
\label{ssec:discussion}
Our two-stage logarithmic approach addresses key challenges in frequency-domain supervision for microscopy image restoration. Our adaptive weighting $w_{\text{rel}}(u,v)$ ensures all frequency components receive attention proportional to their reconstruction difficulty rather than their signal strength. Our focal scaling factor $\alpha$ controls this emphasis: smaller values (e.g., $\alpha=0.5$) compress the weight range for more uniform weights, while larger values (e.g., $\alpha=2$) increase emphasis on challenging frequencies. Since high-frequency regions can contain both fine structural details and noise artifacts, larger $\alpha$ may amplify residual noise in ground-truth images. In practice, we recommend $\alpha$ = 1
as a balanced default, or $\alpha$= 0.5 to prioritize perceptual quality.

%To achieve the aim of balanced frequency loss, we also considered a fully relative error formulation: $\log\frac{|\mathcal{F}_{\text{pred}}| + 1}{|\mathcal{F}_{\text{target}}| + 1}$. However, its gradient $\propto \frac{1}{|\mathcal{F}_{\text{pred}}| + 1}$ vanishes for large amplitudes. Instead, our formulation $D_{\text{log}} = \log(|\Delta| + 1)$ where $\Delta = \mathcal{F}_{\text{pred}} - \mathcal{F}_{\text{target}}$, bases errors on difference magnitude rather than signal amplitude. As shown in Eq.~\ref{ssec:grad}, this yields gradients that scale inversely with the reconstruction error rather than the signal amplitude. This allows high-amplitude, low-frequency components to receive strong gradients when their errors are small. 
Our proposed logarithmic compression decouples loss magnitude from gradient strength: high-amplitude errors contribute more to the total loss but produce proportionally smaller gradients, preventing them from overwhelming optimization and ensuring the network attends to errors across all frequencies.
We compared to alternative frequency-domain losses which face a common challenge: balancing supervision across frequencies with vastly different amplitudes. In FFL \cite{jiang2021}, both weights and errors scale up with raw amplitude, creating squared or cubic dependency on signal strength. This causes the loss to collapse sharply once low frequencies converge, destabilizing training. FSL \cite{fuoli2021} avoids weight tuning but also allows high-amplitudes to dominate, limiting fine-detail supervision for restoration tasks.

\section{Conclusion}
\label{sec:Conclusion}
We introduced a logarithmic focal frequency loss (LFFL) for microscopy image restoration, addressing foreground-background imbalance and spectral bias. LFFL can be easily integrated into GAN-based frameworks without architectural changes. Experiments show strong structural and perceptual gains for deblurring and improvements in pixel fidelity for denoising, demonstrating its effectiveness and practicality for bioimage restoration and related reconstruction tasks.

% Below is an example of how to insert images. Delete the ``\vspace'' line,
% uncomment the preceding line ``\centerline...'' and replace ``imageX.ps''
% with a suitable PostScript file name.
% -------------------------------------------------------------------------
% \begin{figure}[htb]

% \begin{minipage}[b]{1.0\linewidth}
%   \centering
%   \centerline{\includegraphics[width=8.5cm]{example-image}}
% %  \vspace{2.0cm}
%   \centerline{(a) Result 1}\medskip
% \end{minipage}
% %
% \begin{minipage}[b]{.48\linewidth}
%   \centering
%   \centerline{\includegraphics[width=4.0cm]{example-image}}
% %  \vspace{1.5cm}
%   \centerline{(b) Results 3}\medskip
% \end{minipage}
% \hfill
% \begin{minipage}[b]{0.48\linewidth}
%   \centering
%   \centerline{\includegraphics[width=4.0cm]{example-image}}
% %  \vspace{1.5cm}
%   \centerline{(c) Result 4}\medskip
% \end{minipage}
% %
% \caption{Example of placing a figure with experimental results.}
% \label{fig:res}
% %
% \end{figure}

% \vfill
% \pagebreak

\section{Compliance with ethical standards}
\label{sec:ethics}

This study utilized anonymized patient-derived cells provided by clinical collaborators at Aix Marseille University and Marseille Hospital (AP-HM), France, collected under approved institutional protocols. Zebrafish embryo data are made available in open access (MIT license)~\cite{zhang2018poisson} with no ethical approval requirement.

% IEEE-ISBI supports the standard requirements on the use of animal and
% human subjects for scientific and biomedical research. For all IEEE
% ISBI papers reporting data from studies involving human and/or
% animal subjects, formal review and approval, or formal review and
% waiver, by an appropriate institutional review board or ethics
% committee is required and should be stated in the papers. For those
% investigators whose Institutions do not have formal ethics review
% committees, the principles  outlined in the Helsinki Declaration of
% 1975, as revised in 2000, should be followed.

% Reporting on compliance with ethical standards is required
% (irrespective of whether ethical approval was needed for the study) in
% the paper. Authors are responsible for correctness of the statements
% provided in the manuscript. Examples of appropriate statements
% include:
% \begin{itemize}
%   \item ``This is a numerical simulation study for which no ethical
%     approval was required.'' 
%   \item ``This research study was conducted retrospectively using
%     human subject data made available in open access by (Source
%     information). Ethical approval was not required as confirmed by
%     the license attached with the open access data.''
%     \item ``This study was performed in line with the principles of
%       the Declaration of Helsinki. Approval was granted by the Ethics
%       Committee of University B (Date.../No. ...).''
% \end{itemize}

\section{Acknowledgments}
\label{sec:acknowledgments}

This work is funded in part by an endowment in Cardiovascular Bioengineering from the AXA Research Fund (to A.I.B.) and a doctoral fellowship from Institut Polytechnique de Paris (to X.Z.). The authors thank Prof. Catherine Badens and Dr. Camille Desgrouas for providing cell samples.

\bibliographystyle{IEEEbib}
\bibliography{strings,refs}

\end{document}